\documentclass[12pt,a4paper]{panl}
\usepackage{cite}
\usepackage{wrapfig}
\usepackage{graphicx}
\usepackage{amssymb}
\usepackage{amsfonts}
\usepackage{amsmath}
\usepackage{longtable}
\usepackage{rotating}
\usepackage{lscape}
\usepackage{epsfig}
\usepackage{multirow}

\usepackage{geometry}
 \usepackage{setspace}
\usepackage{combelow} 
\DeclareMathOperator{\Tr}{Tr}

\usepackage[dvipsnames]{xcolor}

\originalTeX
\begin{document}

\title{\bf 
On the angular momentum and free energy\\
of rotating gluon plasma\\[3mm]
}
\maketitle
\authors{
V.\,Braguta$^{\,a}$, 
M.\,Chernodub$^{\,b,c}$, 
E.\,Eremeev$^{\,d,}$\footnote{E-mail: eremeev.ei@phystech.edu}, 
I.\,Kudrov$^{\,e}$, 
A.\,Roenko$^{\,a}$, 
D.\,Sychev$^{\,a,d}$
}
\setcounter{footnote}{0}
\from{$^{a}$\,\footnotesize Bogoliubov Laboratory of Theoretical Physics, Joint Institute for Nuclear Research, Dubna, 141980, Russia}
\from{$^{b}$\,\footnotesize  Institut Denis Poisson CNRS UMR 7013, Universit\'e de Tours, Universit\'e d'Orl\'eans, Tours, 37200, France}
\from{$^{c}$\,\footnotesize  Department of Physics, West University of Timi\cb{s}oara, Timi\cb{s}oara, 300223, Romania}
\from{$^{d}$\,\footnotesize  Moscow Institute of Physics and Technology, Dolgoprudny, 141700, Russia}
\from{$^{e}$\,\footnotesize  Institute for High Energy Physics, NRC ``Kurchatov Institute'', Protvino, 142281, Russia}

\begin{abstract}
\begin{spacing}{0.96}
\vskip 2mm
We study the free energy and the angular momentum of rotating hot gluon matter using first-principle numerical simulations of the $\textrm{SU}(3)$ lattice Yang-Mills theory. We calculate the specific moment of inertia and the specific deformation of the gluon matter as, respectively, the leading and next-to-leading terms in a series in angular velocity over a broad range of temperatures and various spatial boundary conditions. We show that the specific deformation, similarly to the moment of inertia, takes negative values in a phenomenologically interesting region of temperatures above the phase transition and turns positive at higher temperatures.
\end{spacing}
\end{abstract}



\section*{\bf \large Introduction}
\vskip 2mm
\label{sec:intro}
Modern experiments on heavy-ion collisions provide us with unique access to the effects emerging in quark-gluon plasma (QGP) under various extreme conditions including high temperatures, strong magnetic fields and baryon densities. Since the QGP is described by Quantum Chromodynamics (QCD), these experiments allow us to test the theory of strong interaction in different regimes, thereby improving our understanding of its nonperturbative properties.

One of the extreme environments is given by a rapid rotation, which has attracted significant interest in the last decade. The STAR collaboration~\cite{STAR:2017ckg} measured the global polarization of $\Lambda, \bar \Lambda$-hyperons and estimated an average vorticity as $\omega \sim 10^{-22}~\textrm{s}^{-1} \sim 10~\textrm{MeV}$. Hydrodynamical simulations predict that vorticity may reach even larger values, $\omega \sim 20{-}40$~MeV~\cite{Jiang:2016woz}, which may significantly affect the thermodynamic properties of QGP.

The theoretical analysis of a vortical system is drastically simplified under the assumption that the vorticity is constant throughout the whole volume, implying that the system rotates with a uniform angular velocity $\Omega$ similar to a rigid body. In this case, friction between cylindrical layers is absent and the system resides in the thermodynamic equilibrium.

Various aspects of the rotating QGP have been studied in different approaches~\cite{Chernodub:2016kxh, Jiang:2016wvv, Chernodub:2017ref, Wang:2018sur, Chen:2020ath,  Golubtsova:2022ldm, Chen:2022smf, Jiang:2023zzu, Sun:2023kuu, Chen:2023cjt, Zhao:2022uxc, Yadav:2022qcl, Braga:2022yfe,Sadooghi:2021upd, Fujimoto:2021xix, Zhang:2020hha, Chen:2022mhf, Singha:2024tpo, Morales-Tejera:2025qvh, Dwibedi:2025boz, Jiang:2021izj, Mameda:2023sst, Chen:2024jet, Sun:2024anu, Chernodub:2020qah, Braga:2023qej, Nunes:2024hzy, Fukushima:2025hmh, Wang:2025mmv, Dey:2025wqw,Kuboniwa:2025vpg, Braga:2025eiz, Chen:2024tkr, Jiang:2024zsw, Yamamoto:2013zwa, Braguta:2020biu, Braguta:2021jgn, Braguta:2022str, Yang:2023vsw, Braguta:2023yjn, Braguta:2023kwl, Braguta:2023tqz, Braguta:2023iyx, Braguta:2024zpi}.
Lattice calculations predict exotic properties of rotating QGP, including a spatially inhomogeneous mixed phase~\cite{Braguta:2023iyx, Braguta:2024zpi} and a negative moment of inertia~\cite{Braguta:2023tqz, Braguta:2023yjn, Braguta:2023kwl}, which may be a manifestation of negative spin-vortical coupling for gluons at temperatures close to the phase transition (``the negative Barnett effect'')~\cite{Braguta:2023tqz}.
These non-trivial effects may also be related to the anisotropy in the gluonic action developed at the curved co-rotating background~\cite{Braguta:2024zpi}.

In this paper, we study the mechanical properties of the rotating hot gluon matter focusing on its free energy and angular momentum. 
We calculate two leading terms in a series of these quantities over the angular velocity thus extending our previous study~\cite{Braguta:2023tqz, Braguta:2023yjn, Braguta:2023kwl}, where only the leading term has been addressed.

\section*{\bf \large Rotating gluodynamics on the lattice}
\vskip 2mm

\label{sec:rotation}
We consider the system in thermal equilibrium in the reference frame rotating together with the system around the $z$-axis. 
To determine the coordinates in this corotating non-inertial frame, $x^\alpha = (t,x,y,z) = ({t}, {r} \cos{{\varphi}}, {r} \sin{{\varphi}}, {z})$, one uses the following transformation, $\varphi = \bar \varphi - \Omega t$, $t = \bar t$, $z = \bar z$, where $x_{\text{lab}}^\alpha \equiv \bar x^\alpha = (\bar{t}, \bar{r} \cos{\bar{\varphi}}, \bar{r} \sin{\bar{\varphi}}, \bar{z})$ are the coordinates in the inertial laboratory frame. 
This transformation leads us to the curvilinear metric,
\begin{equation}\label{eq:interval_Mink}
    ds^2 = g_{\mu\nu} dx^\mu dx^\nu = (1 - r^2 \Omega^2) dt^2 + 2y\Omega\, dt\,dx - 2x\Omega\, dt\,dy - dx^2 - dy^2 - dz^2\,,
\end{equation}
which reduces the effect of rotation to that of an external gravitational field.
Having in mind the numerical simulations that are performed in the Euclidean spacetime, it is convenient to make a Wick rotation and consider the Euclidean metric:
\begin{equation}\label{eq:interval_Eucl}
    ds^2 = g^E_{\mu\nu} dx^\mu dx^\nu = (1 + r^2 \Omega_I^2) d\tau^2 - 2y\Omega_I\, d\tau\,dx + 2x\Omega_I\, d\tau\,dy + dx^2 + dy^2 + dz^2\,,
\end{equation}
where we introduced the imaginary angular velocity $\Omega_I = \partial \bar \varphi / \partial \tau = - i \partial \bar \varphi / \partial t = -i \Omega$. 

The partition function of gluons in the curved background~\eqref{eq:interval_Mink} can be written as a path integral over the gluon fields,
\begin{equation}\label{eq:partition_function}
    \mathcal{Z} = \Tr \Bigl[ e^{-\hat H/T} \Bigr] =
    \int\! \mathcal{D}A\, e^{-S_G[A]}\,,
\end{equation}
where $\hat H$ is the Hamiltonian of the system in the rotating frame.
The Euclidean action, $S_G$, is given by the following general expression,
\begin{equation}\label{eq:Sg_def_E}
    S_G = \frac{1}{4 g_{YM}^2}\int d^4 x \sqrt{g_E}\,  g^{\mu\alpha}_E g^{\nu\beta}_E F_{\mu\nu}^a F_{\alpha\beta}^a\,,
\end{equation}
where $F_{\mu\nu}^a$ is the gluon field-strength tensor and the inverse temperature $1/T$ sets the size of the system in a compact Euclidean time $\tau = it$.

As a result, the action of rotating gluons has a structure of quadratic polynomial in the angular velocity,
\begin{equation}\label{eq:Sg_structure_E}
	S_{G} = S_0 + \Omega_I S_1 + \Omega_I^2 S_2\,,
\end{equation}
where $S_0$ is the action of non-rotating gluodynamics, operator $S_1$ is related to the angular momentum at vanishing rotation, given by a product of chromoelectric and chromomagnetic fields, and operator $S_2$ contains the squares of chromomagnetic fields only. The explicit forms of these operators can be found in Refs.~\cite{Yamamoto:2013zwa,Braguta:2021jgn,Braguta:2024zpi}.

Note that the Euclidean action~\eqref{eq:Sg_structure_E} would be complex-valued in the case of real rotation, which makes it impossible to simulate a real-rotating system directly within the Monte Carlo technique due to the sign problem. In contrast, for an imaginary angular velocity, the sign problem is absent so that we can carry out the lattice simulations for imaginary rotation and then analytically continue the results to real angular velocity.
We consider the system of a finite size in $x$ and $y$ directions thus respecting the causality condition $\Omega r < 1$.

We discretize the action for rotating gluons~\eqref{eq:Sg_structure_E} following Refs.~\cite{Yamamoto:2013zwa, Braguta:2021jgn}. We use the tree-level improved Symanzik action for the $\Omega_I$-independent contributions to the action and chair-clover discretization for the terms that explicitly depend on the angular velocity. To set the lattice spacing, $a$, we use the data for the string tension taken from Ref.~\cite{Beinlich:1997ia}.

The simulations are performed on the lattices of the size $N_t \times N_z \times N_s^2$ ($N_x = N_y = N_s$).
In order to check the sensitivity of our results to the presence of the boundaries in the orthogonal directions $x$ and $y$, we studied the system with different, periodic (PBC) and open (OBC), boundary conditions.
Notice that in our work, the temperature $T = 1/N_t a$ is a global parameter identified with the inverse length of the compactified temporal direction. The Ehrenfest-Tolman law for the metric~\eqref{eq:interval_Eucl} implies that the local temperature depends on the radial coordinate and coincides with $T$ only at the axis of rotation (see a discussion in Ref.~\cite{Braguta:2024zpi}).

\section*{\bf \large Angular momentum and free energy of rotating gluodynamics}
\vskip 2mm
\label{sec:eos}
Free energy of a rotating system can be represented as a series in angular velocity:
\begin{equation} \label{eq:F_series_Omega}
    F(T, R, \Omega) = F_0(T, R) - \frac{F_2(T,R)}{2} \Omega^2  - \frac{F_4(T,R)}{4} \Omega^4 - \dots\,,
\end{equation}
where $F_0 = f_0\, V$ is the free energy at vanishing rotation.

For a system of characteristic radius $R$, the coefficients $F_n$ in the series~\eqref{eq:F_series_Omega} depend on the system size, $F_n = \mathsf{i}_n(T)\, V R^n$, where the specific quantities $\mathsf{i}_n$ are functions of temperature~\cite{Braguta:2023tqz, Braguta:2023yjn, Braguta:2023kwl, Dey:2025wqw}.
These quantities, that determine a mechanical response of the thermodynamic ensemble to the rotation with a constant angular velocity, also determine the angular momentum of the system:
\begin{equation}\label{eq:Jz_Mink}
    J \equiv J_z = - \left(\frac{\partial F}{\partial \Omega}\right)_T = 
    V \Omega R^2 \left( \mathsf{i}_2 + \mathsf{i}_4 (\Omega R)^2 + \mathsf{i}_6 (\Omega R)^4 + \dots \right)\,.
\end{equation}
The leading contribution, $\mathsf{i}_2$, is a specific moment of inertia at vanishing angular velocity, whereas the higher-order coefficients describe a deformation of the system due to the mass-energy redistribution in the rotating system.

For the rotating gluodynamics, the specific moment of inertia was calculated in Refs.~\cite{Braguta:2023tqz, Braguta:2023yjn, Braguta:2023kwl} using two different methods. In this study, we present our first results for the specific deformation coefficient, $\mathsf{i}_4$.

For the thermodynamic ensemble with partition function~\eqref{eq:partition_function}, the imaginary angular momentum, $J_I = i J$, reads as follows:
\begin{equation}\label{eq:Jz_from_lnZ}
    J_I = - \left( \frac{\partial F}{\partial \Omega_I} \right)_T = T \left( \frac{\partial \ln \mathcal{Z}}{\partial \Omega_I} \right)_T
    = - T \langle S_1 + 2 \Omega_I S_2 \rangle\,.
\end{equation}
Note that this expression explicitly depends on the angular velocity because the calculations are performed in the rotating reference frame.
Indeed, one may transform the ``rotating'' gluon fields, $F_{\mu\nu}$, to the ones in the laboratory frame, $\bar F_{\mu\nu}$, using the relation
$\bar F_{\mu\nu} = \frac{\partial x^\alpha}{\partial \bar x^\mu} \frac{\partial x^\beta}{\partial \bar x^\nu} F_{\alpha\beta}$. Then, after performing an inverse Wick transformation to real time, the usual expression for the gluon angular momentum in the laboratory frame can be reconstructed, $J = J_{z,\textrm{lab}} = \int d^3 \bar x \left( \bar x T^{0y}_{\text{lab}} - \bar y T^{0x}_{\text{lab}}\right)$, where $T^{\mu\nu}_{\rm lab}$ is the stress-energy tensor of the gluon field.

On the lattice, the dimensionless total imaginary angular momentum~\eqref{eq:Jz_from_lnZ} is calculated as follows,
\begin{equation}\label{eq:Jz_lat}
    \frac{J_I}{V R T^4} =
    \frac{j_I}{R T^4} 
    = - \dfrac{2 N_t^4}{(N_s - 1)} \Big[ \langle s_1 + 2 \Omega_I s_2 \rangle \Big]^\textrm{sub}\,,
\end{equation}
where $s_1 = T/V\cdot S_1$, $s_2 = T/V\cdot S_2$ and $R = (N_s - 1) a/2$ is the distance from the axis of rotation to the lattice boundary (in the $x,y$-plane, the system has the shape of square $2R\times 2R$).
We subtract a zero-temperature contribution in Eq.~\eqref{eq:Jz_lat}, calculated with the same lattice parameters but on the lattice with a large temporal extension, to avoid divergence in the final result, 
$\left[ \langle s_i \rangle\right]^\textrm{sub} = \langle s_i \rangle_{T} - \langle s_i \rangle_{T=0}= \langle s_i \rangle_{N_t} - \langle s_i \rangle_{N_t^{\infty} = N_z}$.
This approach differs from the method used in Ref.~\cite{Yamamoto:2013zwa}, where only the contribution $s_1$ to the angular momentum was calculated without subtraction. 

Figure~\ref{fig:jI_vs_OmegaI}~(left) shows the total angular momentum~$J_I$ as a function of the velocity of rotation $\Omega_I R$ for different temperatures. The data were obtained on the lattice $5\times 30\times 121^2$ with both types of boundary conditions. One can readily see that while the results for OBC and PBC slightly differ from each other, they demonstrate qualitatively the same behavior as functions of the angular velocity.
\begin{figure}[t]
\begin{center}
\includegraphics[width=0.32\linewidth]{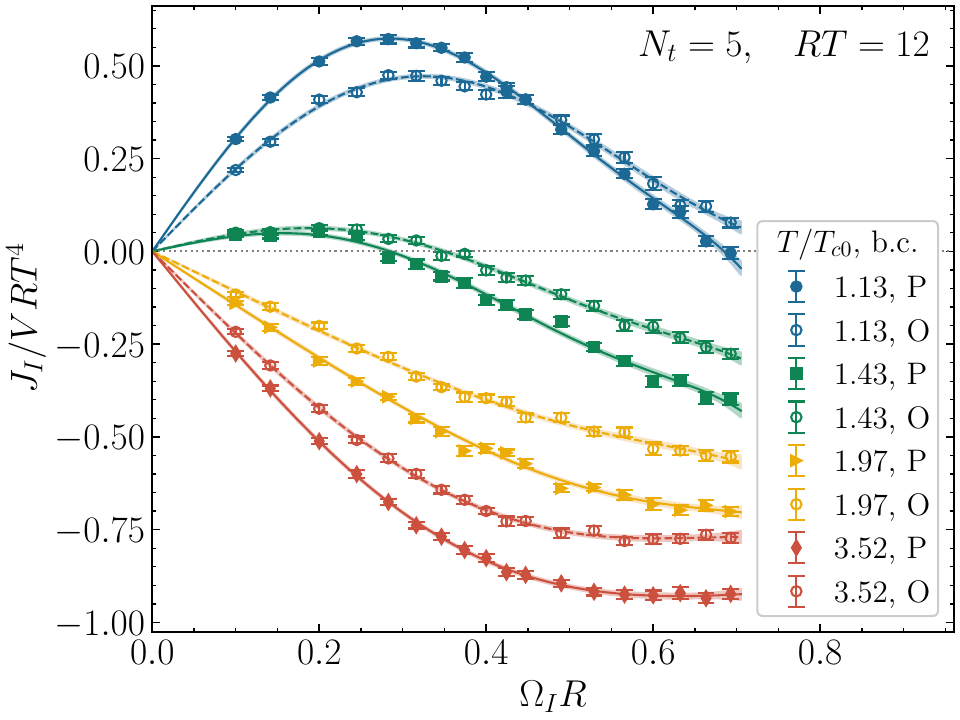}
\hfill
\includegraphics[width=0.32\linewidth]{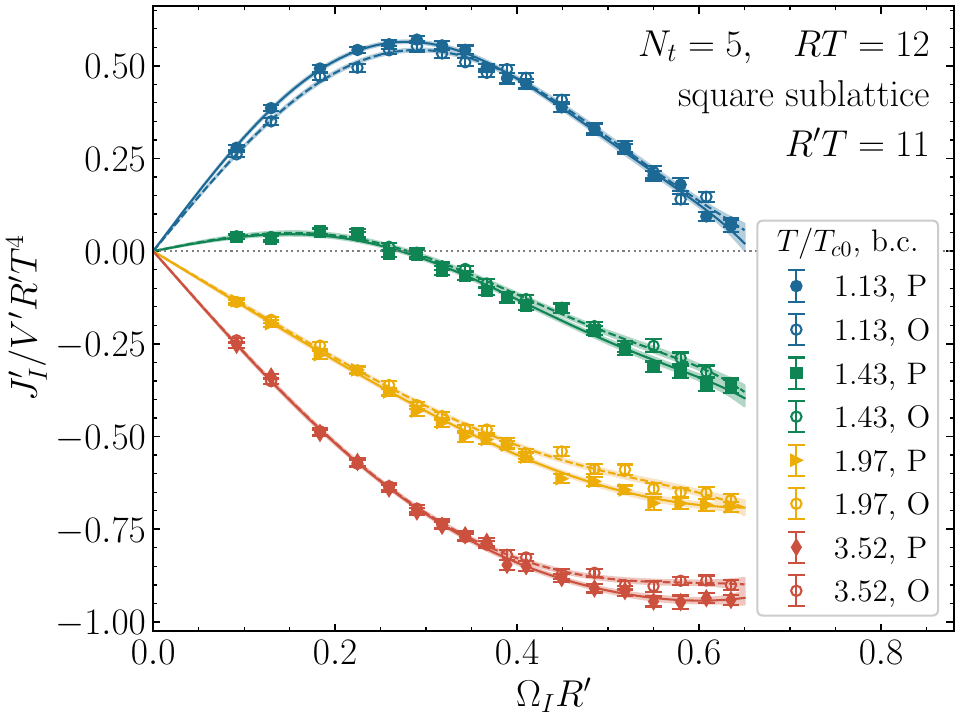}
\hfill
\includegraphics[width=0.32\linewidth]{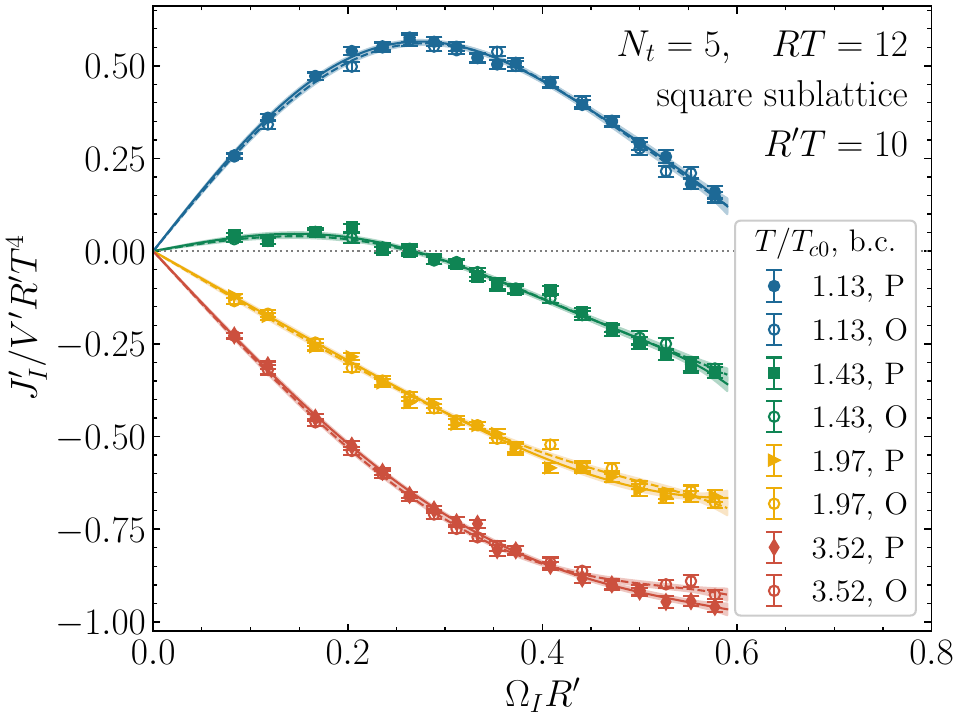}
\vspace{-3mm}
\caption{The imaginary angular momentum as a function of the imaginary angular velocity, calculated on a full lattice (left) and on a square sublattice of size $2R'\times 2R'$ (center, right) with periodic (P) and open (O) boundary conditions for $R'T = 11 $ and $R'T = 10$.
}
\end{center}
\labelf{fig:jI_vs_OmegaI}
\vspace{-5mm}
\end{figure}

To estimate the influence of the boundary conditions on the rotational properties of the system, we also computed the total angular momentum, $J'_I = j'_I V'$, on a square sublattice of size $2R'\times 2R'$, where $R' = R - 1N_t a$ and $R - 2N_t a$. After discarding the outer layer of the width of $(R'-R) = 2N_t a = 2/T$ near the boundary, we find that the results for different boundary conditions are in good agreement with each other (see Fig.~\ref{fig:jI_vs_OmegaI} (right)).
This result confirms our earlier observation that the rotation of the gluonic medium in the bulk is independent of the boundary conditions due to the screening of the boundary effects via the (thermal) gluon mass effects~\cite{Braguta:2021jgn, Braguta:2024zpi}.

We fit the results for total angular momentum by the polynomial function of the imaginary rotational velocity, $v_I = \Omega_I R$:
\begin{equation}\label{eq:jI_series_vI}
    \frac{j_I(T, v_I)}{R T^4} = 
    - \frac{\mathsf{i}_2(T)}{T^4} v_I + \frac{\mathsf{i}_4(T)}{T^4} v_I^3 - \frac{\mathsf{i}_6(T)}{T^4} v_I^5 + \dots \,,
\end{equation}
which is inspired by the representation of the free energy~\eqref{eq:F_series_Omega} for an imaginary rotating system, $\Omega^2 = -\Omega_I^2$.
In our simulations, we consider the domain of the imaginary angular velocities $v_I < 1/\sqrt{2}$, where the causality condition is satisfied. Consequently, the best-fit values $\mathsf{i}_n$ in the fitting function~\eqref{eq:jI_series_vI} correspond to the coefficients in the expansion~\eqref{eq:Jz_Mink} after an analytic continuation, $\Omega^2_I \to -\Omega^2$.

The coefficients $\mathsf{i}_2$ and $\mathsf{i}_4$, calculated on the lattices 
$4 \times 24 \times 97^2$,
$5 \times 30 \times 121^2$,
$6 \times 36 \times 145^2$,
$7 \times 42 \times 169^2$ with PBC, are shown as functions of temperature in Fig.~\ref{fig:in_full_PBC}. We do not present the subsequent coefficients $\mathsf{i}_6$, $\mathsf{i}_8$, etc, due to large statistical uncertainties in their values. The hatched region in Fig.~\ref{fig:in_full_PBC} shows the continuum limit extrapolation, $1/N_t \to 0$.
\begin{figure}
\begin{center}
\includegraphics[width=0.49\linewidth]{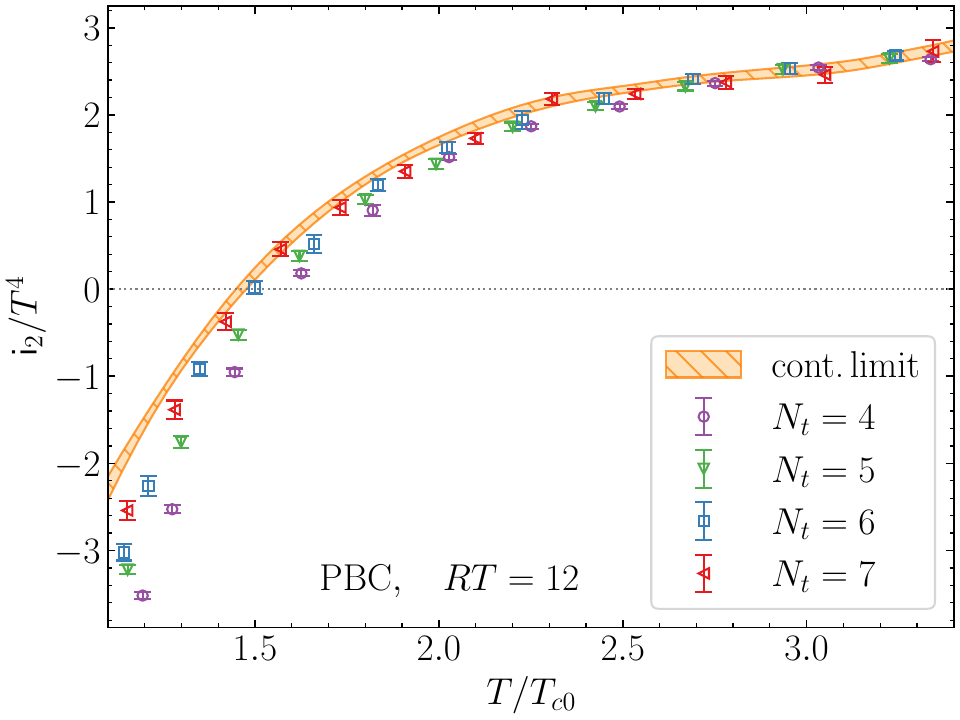}
\hfill
\includegraphics[width=0.49\linewidth]{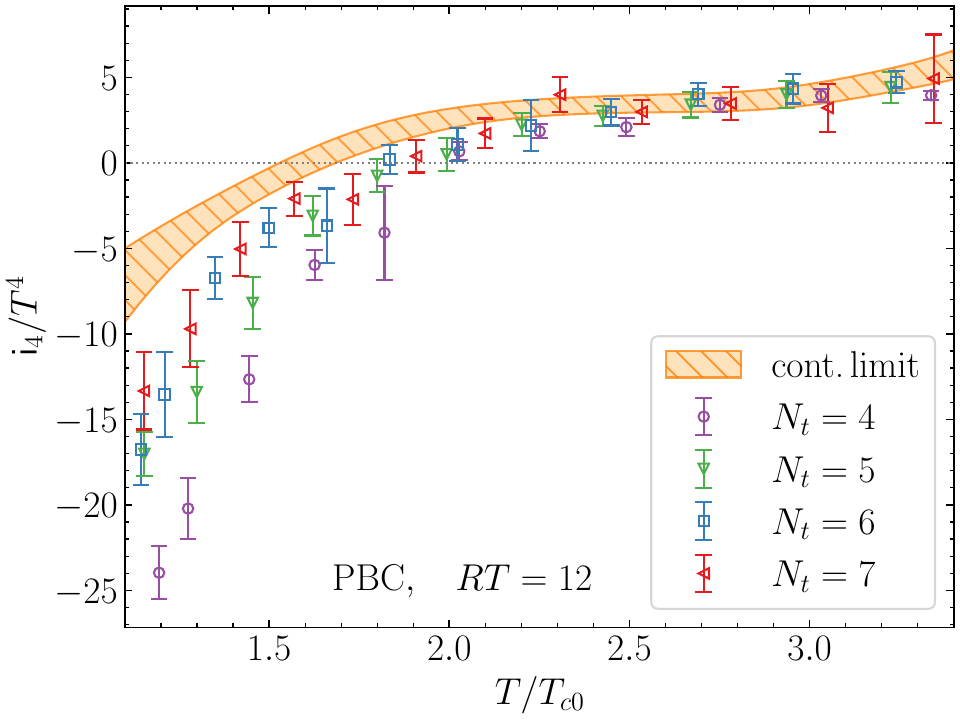}
\vspace{-3mm}
\caption{The specific moment of inertia $\mathsf{i}_2$ and deformation $\mathsf{i}_4$ as a function of temperature computed on the lattices of different sizes with periodic boundary conditions.
}
\end{center}
\labelf{fig:in_full_PBC}
\vspace{-5mm}
\end{figure}

The specific moment of inertia, $\mathsf{i}_2$, takes negative values below the supervortical temperature, $T_s \simeq 1.5 T_{c0}$, where $T_{c0}$ is a critical temperature of the confinement-deconfinement phase transition in non-rotating gluodynamics.
This result, which is in agreement with our previous studies, has been suggested to be a manifestation of the negative Barnett effect~\cite{Braguta:2023tqz}. At high temperatures, the moment of inertia is a positive quantity, in agreement with our intuition originating from classical mechanics.
Asymptotic freedom supports this claim, since at high temperatures -- or, equivalently, at large momentum transfer -- the running QCD coupling becomes small, driving gluon interactions into a weak-coupling, quasi-classical regime.

The specific deformation coefficient, $\mathsf{i}_4$, exhibits a similar behavior in the studied temperature region, $T > 1.1 \,T_{c0}$. The deformation coefficient is positive at high temperatures and becomes negative below a certain fixed value of temperature, $T \sim (1.6-1.8)\, T_{c0}$, which is slightly higher than the supervortical temperature $T_s \simeq 1.5 T_{c0}$.

Note that in classical mechanics, positive values of the deformation coefficient indicate a natural redistribution of the mass to the periphery in the rotating system, whereas the negative values correspond to the opposite behavior: as the rotation frequency increases, the mass tends to gather near the axis of the rotating system.
Let us stress that this mechanical property of the gluon plasma does not fit our intuition coming from the classical mechanics that predict an exactly opposite behavior for a rotating system.
This exotic mechanical property can be correlated with the recent observation of Refs.~\cite{Braguta:2023iyx, Braguta:2024zpi} that rotating gluon gas may form an exotic, spatially inhomogeneous mixed phase within a finite temperature interval above $T_{c0}$.
In this mixed phase, the ``heavy'' deconfinement phase is positioned in the center of the real rotating system, whereas the periphery is in the confinement phase. When the angular velocity increases, the boundary between two phases shifts closer to the rotation axis~\cite{Braguta:2023iyx, Braguta:2024zpi}. We suppose that the negative values of the deformation coefficient is a direct manifestation of this mixed phase.

The results for the coefficients $\mathsf{i}_n$ quantitatively depend on the boundary conditions. 
In Fig.~\ref{fig:in_bc}, the continuum limit extrapolation results are presented for lattices with PBC and OBC,  calculated for the full system of size $RT = 12$ and for square sublattices of different size $R'T = 10, 11$.
From these data, it follows that the effects of boundary conditions for the specific moment of inertia may be of the order of {$\sim 20\%$} at high temperatures and it can reach even $\sim 30-40\%$ at $T=1.1T_{c0}$.
\begin{figure}
\begin{center}
\includegraphics[width=0.49\linewidth]{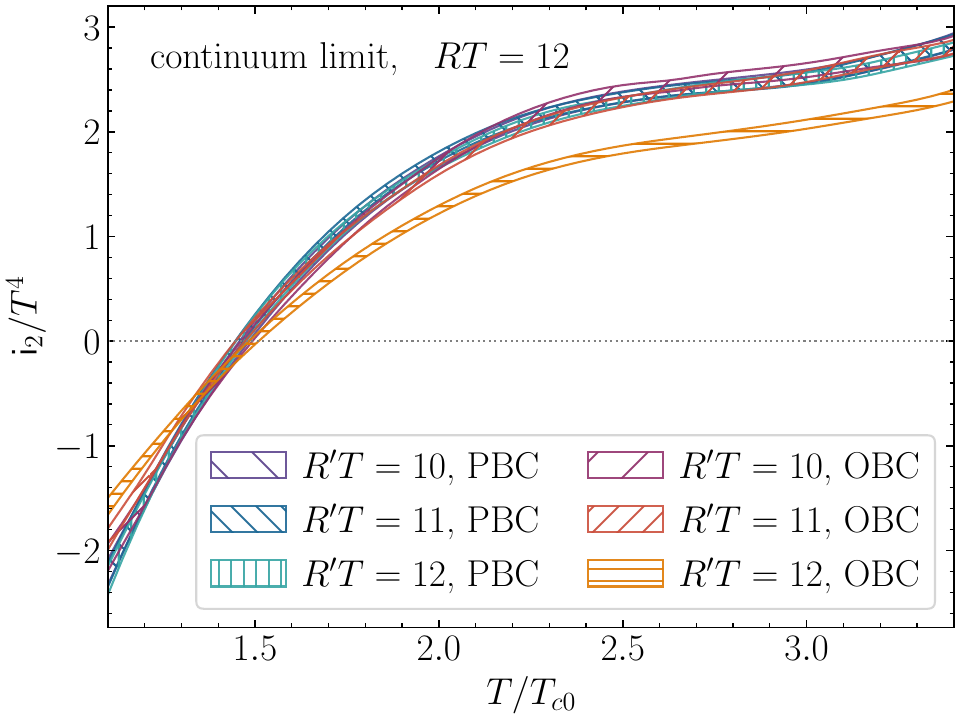}
\hfill
\includegraphics[width=0.49\linewidth]{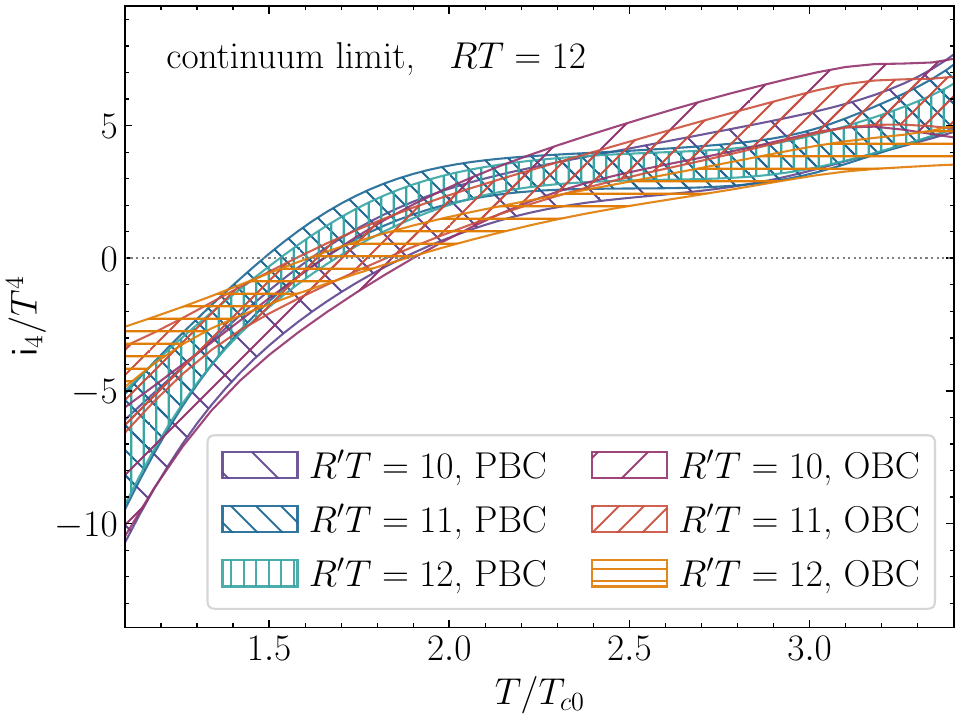}
\vspace{-3mm}
\caption{The specific moment of inertia $\mathsf{i}_2$ and deformation $\mathsf{i}_4$ in the continuum limit as a function of temperature calculated in the lattices with periodic/open boundary conditions and square sublattices of different sizes $2R'$.
}
\end{center}
\labelf{fig:in_bc}
\vspace{-5mm}
\end{figure}

However, the dependence of the results on the boundary condition has a systematic nature which can be easily taken under control. The difference comes as a result of a boundary effect: the results quickly come to agreement after discarding a thin boundary layer in agreement with our results described earlier. The same trend takes place for the deformation coefficient but the effect of discarding the boundaries is not so clearly visible due to the larger uncertainties. Note that the coefficient values $\mathsf{i}_n$ are also affected by the geometry of the system.
These properties of the system will be studied in forthcoming publications devoted to an analysis of angular momentum density.

\section*{\bf \large Conclusions}
\vskip 1mm

Using first-principle lattice simulations, we calculate the angular momentum of rotating $\textrm{SU}(3)$ gluodynamics for different temperatures and angular velocities.
From these data, we extract the specific moment of inertia, $\mathsf{i}_2$, and the specific deformation, $\mathsf{i}_4$, of the rotating gluon system with two different types of boundary conditions in the $x,y$-directions.
The coefficients $\mathsf{i}_n$ determine the free energy of rotating system~\eqref{eq:F_series_Omega} at non-zero angular velocity.
The results for the specific moment of inertia are in agreement with previous studies~\cite{Braguta:2023tqz, Braguta:2023yjn, Braguta:2023kwl}.
It was found that the deformation coefficient is positive at high temperatures, in accordance with classical expectations, but it is negative below $\sim 2 T_{c0}$, which may be a signal of an exotic mixed inhomogeneous phase, predicted for this temperature range in Refs.~\cite{Braguta:2023iyx, Braguta:2024zpi}.

\section*{\bf Acknowledgements}

This work has been carried out using computing resources of the Federal collective usage center Complex for Simulation and Data Processing for Mega-science Facilities at NRC ``Kurchatov Institute'', http://ckp.nrcki.ru/ and the Supercomputer ``Govorun'' of the Joint Institute for Nuclear Research. 

\section*{\bf Funding}
The work of VVB, IEK, AAR, and DAS, which consisted of the lattice calculation of the observables used in the paper,  was supported by the Russian Science Foundation (project no. 23-12-00072). The work of MNC was funded by the EU’s NextGenerationEU instrument through the National Recovery and Resilience Plan of Romania - Pillar III-C9-I8, managed by the Ministry of Research, Innovation and Digitization, within the project entitled ``Facets of Rotating Quark-Gluon Plasma'' (FORQ), contract no. 760079/23.05.2023 code CF 103/15.11.2022.

\section*{\bf Conflict of interest}
The authors declare that they have no conflicts of interest.

\bibliographystyle{pepan}
\bibliography{moment}

\begin{thebibliography}{10}
\def\selectlanguageifdefined#1{
\expandafter\ifx\csname date#1\endcsname\relax
\else\selectlanguage{#1}\fi}
\providecommand*{\href}[2]{{\small #2}}
\providecommand*{\url}[1]{{\small #1}}
\providecommand*{\BibUrl}[1]{\url{#1}}
\providecommand{\BibAnnote}[1]{}
\providecommand*{\BibEmph}[1]{\emph{#1}}
\ProvideTextCommandDefault{\cyrdash}{\hbox to.8em{--\hss--}}
\providecommand*{\BibDash}{}

\bibitem{STAR:2017ckg}
\selectlanguageifdefined{english}
\BibEmph{Adamczyk L. et~al.} [STAR Collaboration] {Global $\Lambda$ hyperon polarization in nuclear collisions: evidence for the most vortical fluid}~// \href{http://dx.doi.org/10.1038/nature23004}{Nature}. \BibDash
\newblock 2017. \BibDash
\newblock V. 548. \BibDash
\newblock P.~62--65. \BibDash
\newblock arXiv:1701.06657.

\bibitem{Jiang:2016woz}
\selectlanguageifdefined{english}
\BibEmph{Jiang Y., Lin Z.W., Liao J.} {Rotating quark-gluon plasma in relativistic heavy ion collisions}~// \href{http://dx.doi.org/10.1103/PhysRevC.94.044910}{Phys. Rev. C}. \BibDash
\newblock 2016. \BibDash
\newblock V.~94, no.~4. \BibDash
\newblock P.~044910. \BibDash
\newblock [Erratum: Phys.Rev.C 95, 049904 (2017)] arXiv:1602.06580.

\bibitem{Chernodub:2016kxh}
\selectlanguageifdefined{english}
\BibEmph{Chernodub M.N., Gongyo S.} {Interacting fermions in rotation: chiral symmetry restoration, moment of inertia and thermodynamics}~// \href{http://dx.doi.org/10.1007/JHEP01(2017)136}{JHEP}. \BibDash
\newblock 2017. \BibDash
\newblock V.~01. \BibDash
\newblock P.~136. \BibDash
\newblock arXiv:1611.02598.

\bibitem{Jiang:2016wvv}
\selectlanguageifdefined{english}
\BibEmph{Jiang Y., Liao J.} {Pairing Phase Transitions of Matter under Rotation}~// \href{http://dx.doi.org/10.1103/PhysRevLett.117.192302}{Phys. Rev. Lett.} \BibDash
\newblock 2016. \BibDash
\newblock V. 117, no.~19. \BibDash
\newblock P.~192302. \BibDash
\newblock arXiv:1606.03808.

\bibitem{Chernodub:2017ref}
\selectlanguageifdefined{english}
\BibEmph{Chernodub M.N., Gongyo S.} {Effects of rotation and boundaries on chiral symmetry breaking of relativistic fermions}~// \href{http://dx.doi.org/10.1103/PhysRevD.95.096006}{Phys. Rev. D}. \BibDash
\newblock 2017. \BibDash
\newblock V.~95, no.~9. \BibDash
\newblock P.~096006. \BibDash
\newblock arXiv:1702.08266.

\bibitem{Wang:2018sur}
\selectlanguageifdefined{english}
\BibEmph{Wang X., Wei M., Li Z., Huang M.} {Quark matter under rotation in the NJL model with vector interaction}~// \href{http://dx.doi.org/10.1103/PhysRevD.99.016018}{Phys. Rev. D}. \BibDash
\newblock 2019. \BibDash
\newblock V.~99, no.~1. \BibDash
\newblock P.~016018. \BibDash
\newblock arXiv:1808.01931.

\bibitem{Chen:2020ath}
\selectlanguageifdefined{english}
\BibEmph{Chen X., Zhang L., Li D., Hou D., Huang M.} {Gluodynamics and deconfinement phase transition under rotation from holography}~// \href{http://dx.doi.org/10.1007/JHEP07(2021)132}{JHEP}. \BibDash
\newblock 2021. \BibDash
\newblock V.~07. \BibDash
\newblock P.~132. \BibDash
\newblock arXiv:2010.14478.

\bibitem{Golubtsova:2022ldm}
\selectlanguageifdefined{english}
\BibEmph{Golubtsova A.A., Tsegelnik N.S.} {Probing the holographic model of N=4 SYM rotating quark-gluon plasma}~// \href{http://dx.doi.org/10.1103/PhysRevD.107.106017}{Phys. Rev. D}. \BibDash
\newblock 2023. \BibDash
\newblock V. 107, no.~10. \BibDash
\newblock P.~106017. \BibDash
\newblock arXiv:2211.11722.

\bibitem{Chen:2022smf}
\selectlanguageifdefined{english}
\BibEmph{Chen S., Fukushima K., Shimada Y.} {Perturbative Confinement in Thermal Yang-Mills Theories Induced by Imaginary Angular Velocity}~// \href{http://dx.doi.org/10.1103/PhysRevLett.129.242002}{Phys. Rev. Lett.} \BibDash
\newblock 2022. \BibDash
\newblock V. 129, no.~24. \BibDash
\newblock P.~242002. \BibDash
\newblock arXiv:2207.12665.

\bibitem{Jiang:2023zzu}
\selectlanguageifdefined{english}
\BibEmph{Jiang Y.} {Rotating SU(2) gluon matter and deconfinement at finite temperature}~// \href{http://dx.doi.org/10.1016/j.physletb.2024.138655}{Phys. Lett. B}. \BibDash
\newblock 2024. \BibDash
\newblock V. 853. \BibDash
\newblock P.~138655. \BibDash
\newblock arXiv:2312.06166.

\bibitem{Sun:2023kuu}
\selectlanguageifdefined{english}
\BibEmph{Sun F., Xu K., Huang M.} {Splitting of chiral and deconfinement phase transitions induced by rotation}~// \href{http://dx.doi.org/10.1103/PhysRevD.108.096007}{Phys. Rev. D}. \BibDash
\newblock 2023. \BibDash
\newblock V. 108, no.~9. \BibDash
\newblock P.~096007. \BibDash
\newblock arXiv:2307.14402.

\bibitem{Chen:2023cjt}
\selectlanguageifdefined{english}
\BibEmph{Chen H.L., Zhu Z.B., Huang X.G.} {Quark-meson model under rotation: A functional renormalization group study}~// \href{http://dx.doi.org/10.1103/PhysRevD.108.054006}{Phys. Rev. D}. \BibDash
\newblock 2023. \BibDash
\newblock V. 108, no.~5. \BibDash
\newblock P.~054006. \BibDash
\newblock arXiv:2306.08362.

\bibitem{Zhao:2022uxc}
\selectlanguageifdefined{english}
\BibEmph{Zhao Y.Q., He S., Hou D., Li L., Li Z.} {Phase diagram of holographic thermal dense QCD matter with rotation}~// \href{http://dx.doi.org/10.1007/JHEP04(2023)115}{JHEP}. \BibDash
\newblock 2023. \BibDash
\newblock V.~04. \BibDash
\newblock P.~115. \BibDash
\newblock arXiv:2212.14662.

\bibitem{Yadav:2022qcl}
\selectlanguageifdefined{english}
\BibEmph{Yadav G.} {Deconfinement temperature of rotating QGP at intermediate coupling from M-theory}~// \href{http://dx.doi.org/10.1016/j.physletb.2023.137925}{Phys. Lett. B}. \BibDash
\newblock 2023. \BibDash
\newblock V. 841. \BibDash
\newblock P.~137925. \BibDash
\newblock arXiv:2203.11959.

\bibitem{Braga:2022yfe}
\selectlanguageifdefined{english}
\BibEmph{Braga N.R.F., Faulhaber L.F., Junqueira O.C.} {Confinement-deconfinement temperature for a rotating quark-gluon plasma}~// \href{http://dx.doi.org/10.1103/PhysRevD.105.106003}{Phys. Rev. D}. \BibDash
\newblock 2022. \BibDash
\newblock V. 105, no.~10. \BibDash
\newblock P.~106003. \BibDash
\newblock arXiv:2201.05581.

\bibitem{Sadooghi:2021upd}
\selectlanguageifdefined{english}
\BibEmph{Sadooghi N., Tabatabaee~Mehr S.M.A., Taghinavaz F.} {Inverse magnetorotational catalysis and the phase diagram of a rotating hot and magnetized quark matter}~// \href{http://dx.doi.org/10.1103/PhysRevD.104.116022}{Phys. Rev. D}. \BibDash
\newblock 2021. \BibDash
\newblock V. 104, no.~11. \BibDash
\newblock P.~116022. \BibDash
\newblock arXiv:2108.12760.

\bibitem{Fujimoto:2021xix}
\selectlanguageifdefined{english}
\BibEmph{Fujimoto Y., Fukushima K., Hidaka Y.} {Deconfining Phase Boundary of Rapidly Rotating Hot and Dense Matter and Analysis of Moment of Inertia}~// \href{http://dx.doi.org/10.1016/j.physletb.2021.136184}{Phys. Lett. B}. \BibDash
\newblock 2021. \BibDash
\newblock V. 816. \BibDash
\newblock P.~136184. \BibDash
\newblock arXiv:2101.09173.

\bibitem{Zhang:2020hha}
\selectlanguageifdefined{english}
\BibEmph{Zhang Z., Shi C., He X.T., Luo X., Zong H.S.} {Chiral phase transition inside a rotating cylinder within the Nambu\textendash{}Jona-Lasinio model}~// \href{http://dx.doi.org/10.1103/PhysRevD.102.114023}{Phys. Rev. D}. \BibDash
\newblock 2020. \BibDash
\newblock V. 102, no.~11. \BibDash
\newblock P.~114023. \BibDash
\newblock arXiv:2012.01017.

\bibitem{Chen:2022mhf}
\selectlanguageifdefined{english}
\BibEmph{Chen Y., Li D., Huang M.} {Inhomogeneous chiral condensation under rotation in the holographic QCD}~// \href{http://dx.doi.org/10.1103/PhysRevD.106.106002}{Phys. Rev. D}. \BibDash
\newblock 2022. \BibDash
\newblock V. 106, no.~10. \BibDash
\newblock P.~106002. \BibDash
\newblock arXiv:2208.05668.

\bibitem{Singha:2024tpo}
\selectlanguageifdefined{english}
\BibEmph{Singha P., Ambrus V.E., Chernodub M.N.} {Inhibition of the splitting of the chiral and deconfinement transition due to rotation in QCD: The phase diagram of the linear sigma model coupled to Polyakov loops}~// \href{http://dx.doi.org/10.1103/PhysRevD.110.094053}{Phys. Rev. D}. \BibDash
\newblock 2024. \BibDash
\newblock V. 110, no.~9. \BibDash
\newblock P.~094053. \BibDash
\newblock arXiv:2407.07828.

\bibitem{Morales-Tejera:2025qvh}
\selectlanguageifdefined{english}
\BibEmph{Morales-Tejera S., Ambru{\c{s}} V.E., Chernodub M.N.} {Firewall boundaries and mixed phases of rotating quark matter in linear sigma model}~// \href{http://dx.doi.org/10.1103/4zrn-wgrg}{Phys. Rev. D}. \BibDash
\newblock 2025. \BibDash
\newblock V. 112, no.~5. \BibDash
\newblock P.~054031. \BibDash
\newblock arXiv:2502.19087.

\bibitem{Dwibedi:2025boz}
\selectlanguageifdefined{english}
\BibEmph{Dwibedi A., Sahu D., Dey J., Goswami K., Ghosh S., Sahoo R.} {Shear viscosity and electrical conductivity of rotating quark matter in Nambu--Jona-Lasinio Model}. \BibDash
\newblock 2025. \BibDash 5. \BibDash
\newblock arXiv:2505.03588.

\bibitem{Jiang:2021izj}
\selectlanguageifdefined{english}
\BibEmph{Jiang Y.} {Chiral vortical catalysis}~// \href{http://dx.doi.org/10.1140/epjc/s10052-022-10915-8}{Eur. Phys. J. C}. \BibDash
\newblock 2022. \BibDash
\newblock V.~82, no.~10. \BibDash
\newblock P.~949. \BibDash
\newblock arXiv:2108.09622.

\bibitem{Mameda:2023sst}
\selectlanguageifdefined{english}
\BibEmph{Mameda K., Takizawa K.} {Deconfinement transition in the revolving bag model}~// \href{http://dx.doi.org/10.1016/j.physletb.2023.138317}{Phys. Lett. B}. \BibDash
\newblock 2023. \BibDash
\newblock V. 847. \BibDash
\newblock P.~138317. \BibDash
\newblock arXiv:2308.07310.

\bibitem{Chen:2024jet}
\selectlanguageifdefined{english}
\BibEmph{Chen Y., Chen X., Li D., Huang M.} {Deconfinement and chiral restoration phase transition under rotation from holography in an anisotropic gravitational background}~// \href{http://dx.doi.org/10.1103/PhysRevD.111.046006}{Phys. Rev. D}. \BibDash
\newblock 2025. \BibDash
\newblock V. 111, no.~4. \BibDash
\newblock P.~046006. \BibDash
\newblock arXiv:2405.06386.

\bibitem{Sun:2024anu}
\selectlanguageifdefined{english}
\BibEmph{Sun F., Shao J., Wen R., Xu K., Huang M.} {Chiral phase transition and spin alignment of vector mesons in the polarized-Polyakov-loop Nambu\textendash{}Jona-Lasinio model under rotation}~// \href{http://dx.doi.org/10.1103/PhysRevD.109.116017}{Phys. Rev. D}. \BibDash
\newblock 2024. \BibDash
\newblock V. 109, no.~11. \BibDash
\newblock P.~116017. \BibDash
\newblock arXiv:2402.16595.

\bibitem{Chernodub:2020qah}
\selectlanguageifdefined{english}
\BibEmph{Chernodub M.N.} {Inhomogeneous confining-deconfining phases in rotating plasmas}~// \href{http://dx.doi.org/10.1103/PhysRevD.103.054027}{Phys. Rev. D}. \BibDash
\newblock 2021. \BibDash
\newblock V. 103, no.~5. \BibDash
\newblock P.~054027. \BibDash
\newblock arXiv:2012.04924.

\bibitem{Braga:2023qej}
\selectlanguageifdefined{english}
\BibEmph{Braga N.R.F., Junqueira O.C.} {Inhomogeneity of a rotating quark-gluon plasma from holography}~// \href{http://dx.doi.org/10.1016/j.physletb.2023.138330}{Phys. Lett. B}. \BibDash
\newblock 2024. \BibDash
\newblock V. 848. \BibDash
\newblock P.~138330. \BibDash
\newblock arXiv:2306.08653.

\bibitem{Nunes:2024hzy}
\selectlanguageifdefined{english}
\BibEmph{Nunes R.M., Farias R.L.S., Tavares W.R., Tim{\'o}teo V.S.} {Chiral vortical catalysis constrained by LQCD simulations}~// \href{http://dx.doi.org/10.1103/PhysRevD.111.056026}{Phys. Rev. D}. \BibDash
\newblock 2025. \BibDash
\newblock V. 111, no.~5. \BibDash
\newblock P.~056026. \BibDash
\newblock arXiv:2412.14541.

\bibitem{Fukushima:2025hmh}
\selectlanguageifdefined{english}
\BibEmph{Fukushima K., Shimada Y.} {Imaginary rotating gluonic matter at strong coupling}~// \href{http://dx.doi.org/10.1016/j.physletb.2025.139716}{Phys. Lett. B}. \BibDash
\newblock 2025. \BibDash
\newblock V. 868. \BibDash
\newblock P.~139716. \BibDash
\newblock arXiv:2506.03560.

\bibitem{Wang:2025mmv}
\selectlanguageifdefined{english}
\BibEmph{Wang S., Chen J.X., Hou D., Ren H.C.} {Strong Coupling Expansion of Gluodynamics on a Lattice under Rotation}. \BibDash
\newblock 2025. \BibDash 5. \BibDash
\newblock arXiv:2505.15487.

\bibitem{Dey:2025wqw}
\selectlanguageifdefined{english}
\BibEmph{Dey S.} {Virial theorem for rigidly rotating matter}. \BibDash
\newblock 2025. \BibDash 4. \BibDash
\newblock arXiv:2504.18388.

\bibitem{Kuboniwa:2025vpg}
\selectlanguageifdefined{english}
\BibEmph{Kuboniwa R., Mameda K.} {Finite-Temperature Perturbation Theory of Rotating Scalar Fields}. \BibDash
\newblock 2025. \BibDash 4. \BibDash
\newblock arXiv:2504.04712.

\bibitem{Braga:2025eiz}
\selectlanguageifdefined{english}
\BibEmph{Braga N.R.F., Junqueira O.C.} {Holographic QCD phase diagram for a rotating plasma in the Hawking-Page approach}~// \href{http://dx.doi.org/10.1016/j.physletb.2025.139669}{Phys. Lett. B}. \BibDash
\newblock 2025. \BibDash
\newblock V. 868. \BibDash
\newblock P.~139669. \BibDash
\newblock arXiv:2501.16446.

\bibitem{Chen:2024tkr}
\selectlanguageifdefined{english}
\BibEmph{Chen S., Fukushima K., Shimada Y.} {Inhomogeneous confinement and chiral symmetry breaking induced by imaginary angular velocity}~// \href{http://dx.doi.org/10.1016/j.physletb.2024.139107}{Phys. Lett. B}. \BibDash
\newblock 2024. \BibDash
\newblock V. 859. \BibDash
\newblock P.~139107. \BibDash
\newblock arXiv:2404.00965.

\bibitem{Jiang:2024zsw}
\selectlanguageifdefined{english}
\BibEmph{Jiang Y.} {Inhomogeneous SU(2) gluon matter under rotation}~// \href{http://dx.doi.org/10.1103/PhysRevD.110.054047}{Phys. Rev. D}. \BibDash
\newblock 2024. \BibDash
\newblock V. 110, no.~5. \BibDash
\newblock P.~054047. \BibDash
\newblock arXiv:2406.03311.

\bibitem{Yamamoto:2013zwa}
\selectlanguageifdefined{english}
\BibEmph{Yamamoto A., Hirono Y.} {Lattice QCD in rotating frames}~// \href{http://dx.doi.org/10.1103/PhysRevLett.111.081601}{Phys. Rev. Lett.} \BibDash
\newblock 2013. \BibDash
\newblock V. 111. \BibDash
\newblock P.~081601. \BibDash
\newblock arXiv:1303.6292~[hep-lat].

\bibitem{Braguta:2020biu}
\selectlanguageifdefined{english}
\BibEmph{Braguta V.V., Kotov A.Y., Kuznedelev D.D., Roenko A.A.} {Study of the Confinement/Deconfinement Phase Transition in Rotating Lattice SU(3) Gluodynamics}~// \href{http://dx.doi.org/10.1134/S0021364020130044}{JETP Lett.} \BibDash
\newblock 2020. \BibDash
\newblock V. 112, no.~1. \BibDash
\newblock P.~6--12.

\bibitem{Braguta:2021jgn}
\selectlanguageifdefined{english}
\BibEmph{Braguta V.V., Kotov A.Y., Kuznedelev D.D., Roenko A.A.} {Influence of relativistic rotation on the confinement-deconfinement transition in gluodynamics}~// \href{http://dx.doi.org/10.1103/PhysRevD.103.094515}{Phys. Rev. D}. \BibDash
\newblock 2021. \BibDash
\newblock V. 103, no.~9. \BibDash
\newblock P.~094515. \BibDash
\newblock arXiv:2102.05084.

\bibitem{Braguta:2022str}
\selectlanguageifdefined{english}
\BibEmph{Braguta V.V., Kotov A., Roenko A., Sychev D.} {Thermal phase transitions in rotating QCD with dynamical quarks}~// \href{http://dx.doi.org/10.22323/1.430.0190}{PoS}. \BibDash
\newblock 2023. \BibDash
\newblock V. 430 (LATTICE2022). \BibDash
\newblock P.~190. \BibDash
\newblock arXiv:2212.03224.

\bibitem{Yang:2023vsw}
\selectlanguageifdefined{english}
\BibEmph{Yang J.C., Huang X.G.} {QCD on Rotating Lattice with Staggered Fermions}. \BibDash
\newblock 2023. \BibDash 7. \BibDash
\newblock arXiv:2307.05755.

\bibitem{Braguta:2023yjn}
\selectlanguageifdefined{english}
\BibEmph{Braguta V.V., Chernodub M.N., Roenko A.A., Sychev D.A.} {Negative moment of inertia and rotational instability of gluon plasma}~// \href{http://dx.doi.org/10.1016/j.physletb.2024.138604}{Phys. Lett. B}. \BibDash
\newblock 2024. \BibDash
\newblock V. 852. \BibDash
\newblock P.~138604. \BibDash
\newblock arXiv:2303.03147.

\bibitem{Braguta:2023kwl}
\selectlanguageifdefined{english}
\BibEmph{Braguta V.V., Kudrov I.E., Roenko A.A., Sychev D.A., Chernodub M.N.} {Lattice Study of the Equation of State of a Rotating Gluon Plasma}~// \href{http://dx.doi.org/10.1134/S0021364023600830}{JETP Lett.} \BibDash
\newblock 2023. \BibDash
\newblock V. 117, no.~9. \BibDash
\newblock P.~639--644.

\bibitem{Braguta:2023tqz}
\selectlanguageifdefined{english}
\BibEmph{Braguta V.V., Chernodub M.N., Kudrov I.E., Roenko A.A., Sychev D.A.} {Negative Barnett effect, negative moment of inertia of the gluon plasma, and thermal evaporation of the chromomagnetic condensate}~// \href{http://dx.doi.org/10.1103/PhysRevD.110.014511}{Phys. Rev. D}. \BibDash
\newblock 2024. \BibDash
\newblock V. 110, no.~1. \BibDash
\newblock P.~014511. \BibDash
\newblock arXiv:2310.16036.

\bibitem{Braguta:2023iyx}
\selectlanguageifdefined{english}
\BibEmph{Braguta V.V., Chernodub M.N., Roenko A.A.} {New mixed inhomogeneous phase in vortical gluon plasma: First-principle results from rotating SU(3) lattice gauge theory}~// \href{http://dx.doi.org/10.1016/j.physletb.2024.138783}{Phys. Lett. B}. \BibDash
\newblock 2024. \BibDash
\newblock V. 855. \BibDash
\newblock P.~138783. \BibDash
\newblock arXiv:2312.13994.

\bibitem{Braguta:2024zpi}
\selectlanguageifdefined{english}
\BibEmph{Braguta V.V., Chernodub M.N., Gershtein Y.A., Roenko A.A.} {On the origin of mixed inhomogeneous phase in vortical gluon plasma}~// \href{http://dx.doi.org/10.1007/JHEP09(2025)079}{JHEP}. \BibDash
\newblock 2025. \BibDash
\newblock V.~09. \BibDash
\newblock P.~079. \BibDash
\newblock arXiv:2411.15085.

\bibitem{Beinlich:1997ia}
\selectlanguageifdefined{english}
\BibEmph{Beinlich B., Karsch F., Laermann E., Peikert A.} {String tension and thermodynamics with tree level and tadpole improved actions}~// \href{http://dx.doi.org/10.1007/s100520050326}{Eur. Phys. J. C}. \BibDash
\newblock 1999. \BibDash
\newblock V.~6. \BibDash
\newblock P.~133--140. \BibDash
\newblock arXiv:hep-lat/9707023.

\end{thebibliography}

\end{document}